\documentclass[conference,compsoc]{IEEEtran}
\IEEEoverridecommandlockouts

\usepackage{algorithm}
\usepackage{algpseudocode}
\usepackage{balance}
\usepackage{booktabs}
\usepackage{graphicx}
\usepackage{tabularx}
\usepackage{url}
\usepackage{amssymb}
\usepackage{enumitem}
\usepackage{xspace}
\usepackage{amsmath}
\usepackage{multirow}
\usepackage{acronym}
\usepackage[hidelinks]{hyperref}

\newacro{CDN}{Content Delivery Network}
\newacro{CSRF}{Cross-Site Request Forgery}
\newacro{WCD}{Web Cache Deception}
\newacro{HTTP}{Hypertext Transfer Protocol}
\newacro{MIDE}{Method Interchange Detection Engine}

\begin{document}

\date{}

\title{When POST Becomes GET: Investigating HTTP Method Interchangeability and Its Security Implications}
\hypersetup{
  pdftitle={When POST Becomes GET: Investigating HTTP Method Interchangeability and Its Security Implications},
  pdfauthor={Matteo Golinelli, Seif Errahmane Hannachi, and Bruno Crispo}
}

\author{\IEEEauthorblockN{Matteo Golinelli}
\IEEEauthorblockA{University of Trento\\
matteo.golinelli@unitn.it}
\and
\IEEEauthorblockN{Seif Errahmane Hannachi}
\IEEEauthorblockA{Higher School of Computer Science of Sidi Bel Abbes\\
s.hannachi@esi-sba.dz}
\and
\IEEEauthorblockN{Bruno Crispo}
\IEEEauthorblockA{University of Trento\\
bruno.crispo@unitn.it}}

\maketitle

\begin{abstract}

HTTP method semantics are a common boundary between safe retrieval and state-changing operations; however, deployed web applications do not always enforce this boundary consistently. In this paper, we study HTTP method interchangeability, a behavior where a POST request can be rewritten as a GET request to the same endpoint, with POST body parameters moved into the query string, while preserving an equivalent observed response. We present \ac{MIDE}, an automated pipeline that crawls web applications in a browser, records POST requests generated by page loads and user-interface interactions, converts supported POST bodies into GET candidates, and compares replayed responses with the original POST response. To avoid counting generic or parameter-independent responses as interchangeable, \ac{MIDE} also sends a no-query GET baseline. We treat as high-confidence evidence of interchangeability only the cases where the converted GET matches the POST, while the baseline does not. We perform a large-scale unauthenticated experiment on the Tranco top 10k. We find high-confidence interchangeable requests on 407 of 10,000 tested sites (4.1\%), 407 of 3,200 sites with captured POSTs (12.7\%), and 407 of 2,734 sites with replayed POSTs (14.9\%). A targeted \ac{CSRF} token-replay check finds cases where invalid tokens are rejected for POST but accepted for the equivalent GET. We also observe \ac{WCD} conditions and cacheable GET responses for interchangeable request instances. These results demonstrate that method interchangeability is observable in real web applications and can expose method-scoped defense gaps.

\end{abstract}

\acresetall


\section{Introduction}
\label{sec:introduction}

\ac{HTTP} methods play an important role in the web security model. GET is intended for safe retrieval of resources, while POST is used for operations that may submit data or change server-side state~\cite{rfc7231}. Web applications, browser behavior, caching systems, and server-side defenses often build on this distinction. For example, SameSite cookie behavior treats safe and unsafe methods differently, and caching layers generally only cache GET responses. When an application accepts the same operation through both methods, these assumptions can become weaker than the application developer intended.

In this paper, we study \emph{HTTP method interchangeability}, the ability of a server to process requests with different HTTP methods in a semantically equivalent way. We focus on a specific and measurable form of the behavior, converting browser-observed POST requests into GET requests to the same endpoint, moving POST body parameters into the query string, and comparing the resulting GET response with the original POST response.

This behavior is not inherently a vulnerability: a read-only endpoint may intentionally accept both methods, and an equivalent response alone does not prove state change or data leakage. Nevertheless, it is security-relevant because it exposes alternative request forms that may bypass assumptions made by defenses, caches, routing middleware, or application code.

Several common development patterns make this behavior plausible. Some frameworks merge query and body parameters into a single interface, while others require explicit separation. Some routing models restrict handlers to a single method by default, while others leave scripts or handlers reachable through multiple methods unless the application checks the request method explicitly. API technologies also differ: GraphQL and SPARQL intentionally allow GET and POST for read-only operations, while other protocols bind semantics more tightly to specific methods.

Motivated by this observation, we develop \ac{MIDE}, a browser-based collection and replay pipeline for measuring method interchangeability at scale.

Detecting method interchangeability at scale presents a key challenge: a converted GET may match the POST either because the endpoint accepts the rewritten parameters or because it ignores parameters and returns a generic response. We therefore use a baseline-controlled comparison. \ac{MIDE} crawls sites with a real browser, records POST requests from page loads, forms, and selected button interactions, converts supported bodies into GET query strings, and replays both the converted GET and a no-query GET baseline. We count high-confidence interchangeability only when the converted GET is equivalent to the POST and the baseline is not.

We evaluate this methodology on an unauthenticated Tranco top 10k large-scale experiment. To the best of our knowledge, this is the first large-scale measurement of HTTP method interchangeability in the wild. Our crawler captured POST requests on 3,200 sites, and 2,734 sites had at least one POST request that could be converted and replayed as a GET. We find method interchangeability on 407 of 10,000 tested sites (4.1\%), 407 of 3,200 sites with captured POSTs (12.7\%), and 407 of 2,734 sites with replayed POSTs (14.9\%). This affects 646 of 8,291 replayed POST routes (7.8\%) and 1,228 of 14,244 replayed POST requests (8.6\%). We further analyze \ac{CSRF} token handling for 115 token-bearing requests and find 7 cases across 2 sites where invalid tokens are rejected for POST but accepted for the equivalent GET. Finally, we test the 1,228 high-confidence interchangeable request instances for \ac{WCD} and confirm cacheable dynamic responses on 56 of them across 9 of 407 tested sites.

To summarize, we make the following contributions:

\begin{itemize}
    \item We analyze framework and API-technology patterns that enable or constrain method interchangeability, including unified parameter access, routing constraints, and array-parameter parsing. We show that popular technologies exhibit behaviors that can lead to method interchangeability.
    \item We present \acf{MIDE}, a browser-based collection and replay pipeline with a baseline-controlled comparison to avoid counting generic responses as interchangeable, and we release its code and configuration as open source.
    \item We perform the first large-scale measurement of HTTP method interchangeability in the wild and find high-confidence interchangeable requests on 4.1\% of all tested sites, 12.7\% of sites with captured POSTs, and 14.9\% of sites with replayed POSTs.
    \item We analyze method-scoped \acl{CSRF} token behavior and identify cases where the POST form rejects invalid tokens while the equivalent GET form accepts them.
    \item We test high-confidence interchangeable GET URLs for \acl{WCD} and confirm that method interchangeability can make dynamic responses cacheable under static-looking URLs.
\end{itemize}

\subsection{Research Questions}

Our study answers the following research questions.

\begin{enumerate}[label=RQ\arabic{enumi}]
    \item What are the common patterns and behaviors in popular web frameworks and API technologies that can lead to HTTP method interchangeability?
    \label{Q1}
    \item Can we reliably detect HTTP method interchangeability at scale in an automated way?
    \label{Q2}
    \item How prevalent is HTTP method interchangeability in real-world web applications?
    \label{Q3}
    \item What are the security implications of HTTP method interchangeability, particularly in relation to defenses that rely on method semantics?
    \label{Q4}
\end{enumerate}

\subsubsection*{Availability}

We release the \ac{MIDE} code and the configuration used in our experiments as open source at \url{https://github.com/Golim/method-interchange}.

\section{Background}
\label{sec:background}

In this section, we provide background on HTTP method semantics, the concept of method interchangeability, and its implications for web security, particularly in the context of \ac{CSRF} and \ac{WCD} attacks.

\subsection{Method Interchangeability}

\ac{HTTP} defines a set of request methods that indicate the desired action to be performed on a given resource. Among these, GET and POST are the most widely used in web applications. According to the HTTP specification, GET is intended for safe and idempotent operations, meaning it should not cause any state change on the server, while POST is designed for operations that may create or modify server-side state~\cite{rfc7231}. In practice, however, these semantic distinctions are not strictly enforced by many web applications, leading to inconsistencies between intended and actual behavior.

Modern web development frameworks abstract much of the underlying HTTP handling, often routing requests based primarily on URL paths rather than strictly enforcing method semantics. As a result, developers may implement endpoints that accept multiple HTTP methods interchangeably. This can occur due to shared handler functions, permissive routing configurations, or middleware that does not differentiate between methods. Additionally, reverse proxies, load balancers, and caching layers may further obscure method distinctions by normalizing or rewriting requests, contributing to inconsistent behavior across deployment environments. For example, the \texttt{proxy\_method} directive in Nginx can be configured to change the HTTP method of incoming requests when forwarding them upstream to the origin server.

From a security perspective, many web defenses rely on assumptions about HTTP method semantics. Cross-Site Request Forgery (CSRF) protections, for example, are typically designed under the assumption that state-changing operations are not accessible via GET requests. Browsers facilitate this model by allowing cross-origin GET requests to be triggered easily (e.g., via image tags or links), while restricting the ability to issue arbitrary cross-origin POST requests without user interaction or preflight checks. Mechanisms such as SameSite cookies and CSRF tokens are therefore primarily applied to non-idempotent requests. Specifically, cookies with \texttt{SameSite=Lax} attribute are not sent with cross-site POST requests, while they are sent with cross-site GET requests~\cite{ietf-httpbis-rfc6265bis-22}. Similarly, \ac{CSRF} tokens are often embedded in forms that submit via POST, and may not be validated for GET requests. If a server allows state-changing operations via GET, these defenses may be bypassed or severely weakened.

Caching behavior introduces additional complexity. RFC 7231 specifies that GET responses are cacheable by default, provided they meet certain criteria, while POST responses are not, unless they include explicit freshness information~\cite{rfc7231}. If a server responds identically to both POST and GET requests for a given endpoint, and the GET response is cacheable, sensitive, or user-specific data may be stored and served from caches in unintended ways. This discrepancy between application logic and caching policies underlies attacks such as \ac{WCD}, where dynamic content is exposed through cacheable URLs~\cite{wcd_blog,wcd_paper,mirheidari2020cached,mirheidari2022web}.

In real-world applications, the boundary between GET and POST is frequently blurred by layers of abstraction, framework conventions, and developer practices. For example, APIs may accept parameters both in the query string and in the request body, with precedence rules that are not well defined and vary across implementations. Some frameworks automatically merge parameters from multiple sources, leading to ambiguous or inconsistent behavior when the same endpoint is accessed using different methods. In Section~\ref{sec:frameworks} we provide an overview of how different server-side technologies handle HTTP methods, highlighting common patterns that can lead to method interchangeability.

Client-side behavior also plays a role. Browsers automatically generate POST requests for form submissions but may fall back to GET under certain conditions, such as when forms use the default method, or when JavaScript rewrites requests. Additionally, some client-side libraries and frameworks may abstract away HTTP methods, allowing developers to specify actions without explicitly choosing between GET and POST, further contributing to interchangeable behavior.

These discrepancies lead to what is referred to as method interchangeability, the ability of a server to process requests with different HTTP methods in a semantically equivalent way. While this behavior is not inherently a vulnerability, it can expand the attack surface of web applications by enabling alternative interaction patterns that bypass assumptions of security mechanisms and could lead to unexpected security implications. For instance, an endpoint designed to accept POST requests with CSRF protection may also accept equivalent GET requests without enforcing the same protections, enabling exploitation through cross-site request contexts~\cite{owasp-wstg-http-verb-tampering}. Moreover, if the GET response is cacheable, sensitive data may be exposed through caching layers, increasing the risk of information disclosure.

\subsection{Cross-Site Request Forgery}

\ac{CSRF} is a class of attack where an adversary tricks a victim's browser into performing unintended actions on a web application in which they are authenticated without their intent. The attack typically involves crafting a malicious request that exploits the victim's authenticated session, causing the server to execute actions on behalf of the attacker without the victim's consent. \ac{CSRF} attacks rely on the fact that browsers automatically include credentials, such as cookies or HTTP authentication headers, with cross-site requests to the target domain. If a server does not properly validate the origin of requests or enforce method semantics, it may be vulnerable to \ac{CSRF} attacks, particularly if state-changing operations are accessible via GET requests.

Restricting state-changing operations to POST requests is not sufficient to prevent \ac{CSRF} attacks, since attackers can still craft cross-origin POST requests, such as through hidden forms or JavaScript on malicious sites. Therefore, defenses against \ac{CSRF} typically involve a combination of techniques that validate the legitimacy of requests. The most common server-side defense is the use of \ac{CSRF} tokens. The application associates an unpredictable token with the user's session and requires that token to be submitted with state-changing requests, typically as a hidden form field or request header. A related approach is the double-submit cookie pattern, where the application verifies that a token sent in the request matches a cookie.

Modern \ac{CSRF} defenses are built around the assumption that state-changing operations should be restricted to POST requests. Frameworks typically enforce \ac{CSRF} protections, such as tokens or origin checks, only on unsafe methods (e.g., POST, PUT, DELETE), while treating GET as inherently safe and therefore exempt from protection. This design is also reflected in browser-side mitigations such as SameSite cookies. In the commonly deployed \texttt{SameSite=Lax} mode, cookies are included in cross-site top-level navigations only if the request uses a safe method, primarily GET. Consequently, if a state-changing endpoint is reachable via GET, the browser will include authentication cookies in a cross-site request.

Method interchangeability directly violates these assumptions. If an endpoint designed for POST requests can be invoked equivalently via GET, then the attacker gains access to a request primitive that is easier to trigger cross-site and more likely to bypass both browser-level and server-side defenses. This effectively reduces the barrier for exploitation from requiring a crafted form submission or JavaScript execution to a simple link or embedded resource.

Beyond classical \ac{CSRF}, this behavior is particularly relevant for variants such as login \ac{CSRF} and session fixation. In login \ac{CSRF}, the attacker forces the victim to authenticate as the attacker, establishing a session under the attacker's control~\cite{barth2008robust}. If login or session-binding operations are accessible via GET due to method interchangeability, they become significantly easier to trigger. Similarly, any endpoint that sets or modifies authentication state through interchangeable methods can expose subtle session management flaws that would otherwise be harder to exploit. In session fixation, the application fails to renew the session identifier after authentication, allowing an attacker who knows or controls the pre-authentication identifier to access the resulting authenticated session~\cite{owasp-wstg-session-fixation}.

\subsection{Web Cache Deception}

\ac{WCD} arises from inconsistencies between how origin servers and caching layers interpret and handle HTTP requests. In a typical WCD scenario, an attacker crafts a URL that the origin server treats as a dynamic, user-specific resource, while an intermediary cache interprets it as a static, cacheable object. \ac{WCD} can be defined as the behavior where a cache stores and serves a response that was intended to be dynamic and user-specific. This can happen as a consequence of an attacker-modified URL that confuses the caching layer about the static or dynamic nature of the content, but it can also happen without URL manipulation if the application itself serves dynamic content at cacheable URLs. If the response is cached, subsequent requests to the same URL will return sensitive content to other users~\cite{wcd_blog,wcd_paper}. Even when the origin server sets appropriate cache-control headers, intermediary caches may apply their own heuristics or normalization rules. For example, cache keys may depend on the URL but ignore certain headers or parameters. Previous studies have shown that many caching layers, including CDNs and reverse proxies, are frequently configured to cache GET responses solely based on URL patterns, such as checking for the presence of static file extensions at the end of the path~\cite{mirheidari2020cached,mirheidari2022web}, disregarding the cache control headers and other indicators of cacheability. This is generally done to optimize performance or backwards compatibility.

Method interchange affects \ac{WCD} by enabling attackers to cache responses that would otherwise be non-cacheable if accessed via POST. If an endpoint that serves user-specific content is accessible via both POST and GET, and the GET response is cacheable, an attacker can exploit \ac{WCD} to store and steal sensitive information that was only intended to be accessible through POST requests.

\section{Related Work}
\label{sec:related_works}

To the best of our knowledge, there has been no prior systematic measurement of HTTP method interchangeability as a distinct phenomenon. However, our work builds on several related lines of research in web security. Here, we review the most relevant works.

\subsubsection*{HTTP Method Semantics and Verb Tampering}

HTTP assigns semantics to request methods, including safety and cacheability properties. In particular, GET is specified as a safe method, whereas POST requests ask the target resource to process the enclosed representation according to resource-specific semantics; HTTP caching rules also treat GET and POST differently~\cite{rfc9110,rfc9111}. Security testing practice has long recognized that changing HTTP methods can expose security bugs. The OWASP Web Security Testing Guide describes HTTP verb tampering as testing whether application functionality incorrectly accepts alternative methods, and its HTTP method testing guidance discusses method-based access-control bypasses and method-override headers such as \texttt{X-HTTP-Method-Override}~\cite{owaspVerbTampering,owaspHttpMethods}.

\subsubsection*{Parameter-Source and Parser Inconsistencies}

Our transformation is also related to prior work on HTTP parameter handling. Balduzzi et al. studied HTTP Parameter Pollution (HPP), where attackers exploit differences in how applications, frameworks, and intermediaries parse duplicated or encoded parameters~\cite{balduzzi2011automated}. Their work shows that parameter location and parsing rules are security-relevant, especially when components disagree about precedence or interpretation.

\subsubsection*{\ac{CSRF} and Method-Scoped Defenses}

\ac{CSRF} is a long-studied web vulnerability. Barth et al. proposed robust CSRF defenses and analyzed why cross-site requests remain difficult to distinguish from legitimate user actions~\cite{barth2008robust}. Later, several works have focused on automating CSRF detection. Pellegrino et al. propose a system that uses dynamic analysis and property graphs to identify and validate CSRF vulnerabilities in web applications~\cite{pellegrino2017deemon}, while Calzavara et al. use supervised learning to identify sensitive HTTP requests and enable black-box CSRF testing~\cite{calzavara2019mitch}. Other works have studied browser-side mitigations, such as SameSite cookies, including their deployment, effectiveness, and limitations~\cite{khodayari2022samesite}.

\subsubsection*{\ac{WCD} and Cache Inconsistencies}

\acf{WCD} exploits disagreement between an origin server, which treats a URL as dynamic or user-specific, and a cache, which treats the same URL as static and cacheable. Gil introduced \ac{WCD} as a practical attack pattern~\cite{wcd_blog,wcd_paper}. Mirheidari et al. later performed the first large-scale study of \ac{WCD} in the wild and characterized \ac{WCD} as an instance of path confusion between origins and caches~\cite{mirheidari2020cached}. Their follow-up work expanded \ac{WCD} detection using response-identicality checks and cache status headers heuristics, enabling larger-scale measurements~\cite{mirheidari2022web}. Golinelli et al. studied hidden web caches and showed that cache behavior can be difficult to observe when cache status headers are absent or unreliable~\cite{golinelli2024hidden}.

\subsubsection*{Semantic and Policy Inconsistencies on the Web}

More broadly, recent work has shown that inconsistent interpretation of security mechanisms is a common source of vulnerabilities. Calzavara et al. studied inconsistencies between X-Frame-Options and CSP-based framing control, showing that browser and policy differences can lead to different protection guarantees~\cite{calzavara2020tale}. In a later work, Calzavara et al. formalized inconsistencies in cookie attributes, CSP, and HSTS across pages within the same security boundary and proposed Site Policy to make such inconsistencies explicit~\cite{calzavara2021reining}. Although their setting is different, the underlying lesson is similar: when security depends on semantics being interpreted uniformly across components, underspecification and implementation divergence create exploitable gaps. We view HTTP method interchangeability as another instance of this broader phenomenon.

\section{Method Interchangeability in Popular Web Frameworks}
\label{sec:frameworks}

To understand whether method interchangeability could arise from framework defaults rather than from developer mistakes alone, we selected frameworks to cover a diverse set of server-side designs across seven language and platform ecosystems. In particular, the set includes frameworks with unified or partially unified query and body parameter access as well as frameworks with explicit source separation, and both method-permissive and method-restricted routing models, allowing us to examine the main design patterns that can enable or constrain method interchangeability.

The goal of this analysis is not to identify inherent flaws or vulnerabilities in specific frameworks, but rather to characterize design patterns and conventions that may lead to method interchangeability. In particular, we focus on how frameworks route requests to handlers, how they expose request parameters and data to developers, and how they enforce or relax method-specific constraints by default. First, we determine whether frameworks expose a unified interface for accessing data from the query string and the request body, since such abstractions may make GET and POST requests behave similarly at the application layer. Second, we assess whether method interchange is enabled by default, or instead requires explicit routing decisions by developers.

Our analysis combines documentation review with empirical testing. For each framework, we inspected the official request handling model, focusing on whether it provides a single abstraction over query parameters and form or body parameters. We then implemented test applications to determine whether endpoints intended for POST could be reached via GET with equivalent inputs, and whether this behavior occurred by default or required explicit multi-method route configuration. Additionally, we examined how frameworks parse array-valued query parameters, which impacts the correctness of POST-to-GET replay.

Detailed results on parameter-source abstractions, routing constraints, and array parsing are reported in Appendix~\ref{app:framework-details}. We keep the API-technology comparison here because these protocols define HTTP-method semantics independently of server-side framework defaults.

\subsection{Method Interchangeability in API Interface Technologies}
\label{sec:api-technologies}

Application-layer API technologies can also influence method interchangeability. Query-centric technologies such as GraphQL, SPARQL, and OData define their own operation model and then map it onto HTTP~\cite{GraphQL-Spec,SPARQL-Protocol,OData-Protocol}, while JSON:API, gRPC, XML-RPC, and JSON-RPC impose more method-specific semantics~\cite{JSONAPI-Spec,gRPC-over-h2,XML-RPC-Spec,JSON-RPC-Spec}.

\textbf{GraphQL} is the most relevant case for our investigation. The core GraphQL specification is transport-agnostic~\cite{GraphQL-Spec}, but the GraphQL-over-HTTP draft defines the common HTTP semantics for GraphQL operations. Query operations, which are intended to be read-only, must be accepted via POST and may also be accepted via GET, while mutation operations must not be executed via GET~\cite{GraphQL-over-HTTP}. \textbf{SPARQL} is similar: query operations may use either GET or POST, whereas update operations are POST-only~\cite{SPARQL-Protocol}. \textbf{OData} encodes the same distinction differently: functions must have no side effects and are invoked via GET, whereas actions may have side effects and are invoked via POST~\cite{OData-Protocol}. By contrast, \textbf{JSON:API}, \textbf{gRPC}, and \textbf{XML-RPC} are designed around method-specific semantics~\cite{JSONAPI-Spec,gRPC-over-h2,XML-RPC-Spec}, while \textbf{JSON-RPC} is transport-agnostic, so any GET/POST interchange depends on the implementation rather than on the specification~\cite{JSON-RPC-Spec}.

\subsection{Framework Analysis Summary}

Our analysis shows that method interchangeability depends on two independent dimensions: parameter-source abstraction and route reachability. Environments such as vanilla PHP and specific Spring MVC or ASP.NET Core configurations, combine permissive routing with unified or partially unified input handling and therefore are more prone to interchangeability. Others, such as Laravel, Flask, Rails, Sinatra, and Gin, can approximate the same behavior only after explicit multi-method routing is introduced. Finally, frameworks such as Django, Express, FastAPI, NestJS, Next.js, and SvelteKit keep parameter sources more explicit, even if some of them remain method-permissive at the view level. At the API-technology layer, GraphQL and SPARQL intentionally allow GET/POST duality for read-only operations, while other technologies preserve method-specific semantics by design. These findings answer~\ref{Q1}, identifying the framework and API-technology patterns that make method interchangeability more or less likely. These findings inform the design of detection techniques, which must account for both routing semantics and parameter parsing behavior.

\section{Methodology}
\label{sec:methodology}

In this section, we describe our automated pipeline for detecting HTTP method interchangeability in web applications. The pipeline visits a site, records POST requests generated by page loads and user-interface interactions, rewrites supported POST bodies into GET query strings, and compares the resulting GET responses with the original POST responses. The large-scale measurement in Section~\ref{sec:experiment} uses this pipeline without any form of authentication: each site is visited in an unauthenticated browser context.

\subsection{System Overview}

Our system combines browser automation, request collection, replay, and response-similarity analysis. Crawling is implemented with Playwright, using a Chromium browser context for page navigation and passive network observation. The Tranco run used Playwright 1.58.0, \texttt{playwright-stealth} 2.0.2, \texttt{requests} 2.32.5, and bundled Chromium 145.0.7632.6 in headless mode. We did not configure a custom user agent; captured requests, therefore, used Chromium's default headless user agent. Browser contexts used a 1920$\times$1080 viewport and screen, blocked service workers, and enabled \texttt{ignore\_https\_errors}. The browser is restarted between sites.

The design separates collection from replay. The crawler records POST requests and their observed responses, while \ac{MIDE} converts each captured request into one or more GET candidates. This separation makes the analysis repeatable, where changing the conversion strategy or equivalence threshold does not require re-crawling the site.

\subsection{Crawling and Request Collection}

The goal of the crawling phase is to collect POST requests corresponding to realistic browser activity. These include requests triggered by explicit actions, such as form submission or button clicks, as well as background POST requests issued during page load, such as API calls.

We perform a breadth-first crawl starting from the target URL. The crawler stays within the target domain and its subdomains, filters configured denylisted URLs, and records POST requests with resource types \texttt{xhr}, \texttt{fetch}, \texttt{document}, or \texttt{other}. In our experiments, each crawl was bounded to 10 pages and a depth of 5, and the crawler requested shutdown once 100 POST requests had been captured. Page navigation waited for \texttt{domcontentloaded} with a 30-second timeout, followed by a 2-second settle time for JavaScript-triggered requests; the runner used a 900-second per-site timeout.

The crawler deduplicates requests by URL, content type, and POST body, except for configured patterns where repeated requests are intentionally preserved, namely \texttt{/graphql} and \texttt{/graphiql}. It also drops requests matching potentially destructive or state and session-changing action patterns and requests that carry file-upload signals to avoid generating harmful side effects. This filter is applied as a case-insensitive keyword check over captured POST URL paths, while multipart uploads are detected from \texttt{Content-Disposition} parts containing a filename. This category includes logout and sign-out operations: even without intentional authentication, they can alter application state, invalidate transient or pre-authentication sessions, and disrupt subsequent crawl behavior.

At each visited page, the crawler extracts links for further exploration and interacts with POST forms and selected clickable controls. We configured the crawler to process at most 5 forms per page, touch at most 20 fields per form, and click at most 10 standalone buttons or button-like controls per page. Form and button interaction used a 15-second overall budget per page and a 1-second timeout for individual DOM actions. For forms, it fills controls with benign synthetic values according to simple field-type heuristics, removes client-side \texttt{required} and \texttt{pattern} attributes, skips file inputs, and submits the form only when the action stays within the target site and does not match the configured denylist. For clickable controls outside forms, it avoids controls whose text matches potentially destructive or state-changing action patterns. All automatically generated inputs are syntactically valid and benign; the crawler does not inject attack payloads.

The crawler observes network traffic with Playwright page events. When a matching POST request starts, the crawler records the URL, method, headers, POST body, timestamp, and resource type; when Playwright later reports \texttt{requestfinished} or \texttt{requestfailed}, it attaches the observed response or failure record. During this phase, service workers are blocked, since Playwright notes that these can hide network events from request interception APIs and recommends blocking them when request interception is required~\cite{playwright-browsercontext-route}.

\subsection{Method Interchange Detection Engine (MIDE)}
\label{sec:detection}

After crawling, the captured POST requests are processed by \ac{MIDE}. Its purpose is to determine whether the server accepts a GET request for the same endpoint with parameters moved from the POST body into the query string.

\ac{MIDE} supports three POST body formats because they are the principal browser and API representations whose parameters can be systematically mapped into a GET query string:

\begin{itemize}
    \item \textbf{\texttt{application/x-www-form-urlencoded}.} The converter parses form pairs from the POST body and re-encodes them into the GET query string.
    \item \textbf{\texttt{multipart/form-data}.} The converter extracts normal form fields into query parameters. File-upload parts cannot be represented in a GET URL; when such parts are present, they are omitted, and the conversion is marked as lossy.
    \item \textbf{\texttt{application/json}.} The converter flattens JSON into bracket-notation query parameters up to a maximum depth of 2. In our experiments, arrays are encoded using PHP-style empty brackets and, for selected cases, repeated keys; deeper nested structures fail conversion through depth truncation, while explicit null values are omitted and marked as lossy.
\end{itemize}

The default replay path uses the PHP array style. Replay does not try every conversion mode as a fallback chain. If conversion fails for a POST request, replay is skipped for that request. Requests with missing or unsupported \texttt{Content-Type} values are bucketed as unknown and skipped, and malformed JSON is recorded as a malformed body before replay. JSON is a special case: if the payload is a top-level array, or if a top-level object field contains an array, the system creates two GET candidates, one using PHP-style arrays and one using repeated keys. These candidates are sent serially and are not fallback attempts.

For each successful conversion, \ac{MIDE} sends the converted GET request to the same endpoint. It removes body-specific headers such as \texttt{Content-Type} and \texttt{Content-Length}, while preserving the remaining request headers. The replay engine also sends a baseline GET request to the same URL with no query string. This no-query GET is used as a control to identify cases where the endpoint returns the same response even without the converted parameters. Replays are serialized with a 2-second delay between records and a 30-second request timeout. If a server responds with \texttt{429 Too Many Requests}, the implementation retries once after the response's \texttt{Retry-After} value, or after 60 seconds when that header is missing or unparseable. The main replay stage does not resend the original POST request. It must be noted that in our experiments, the replay stage happens right after the crawling stage to avoid significant time gaps between the original POST and the replayed GETs, which could lead to state changes on the server side that affect response equivalence.

\subsection{Response Equivalence Analysis}

A central challenge in detecting method interchangeability is deciding whether the response to a rewritten GET request is equivalent to the response to the original POST request. Exact matching is too strict for real web applications, since semantically equivalent responses often differ in non-essential ways, such as timestamps, dynamic nonces, randomized identifiers, or personalized counters.

We therefore use a response-equivalence criterion based on the captured POST response and the replayed GET response. First, the HTTP status codes must match. Second, the structural similarity score must be at least $0.95$. For HTML responses, structural similarity compares the sequence of start and end tags while ignoring text content and all tag attributes. For JSON responses, it compares nested key paths while ignoring ordinary values; for other bodies, it falls back to a sequence similarity score. The comparator also vetoes JSON responses that disagree on common application-level outcome fields, such as status or success indicators. We compare response bodies as captured, without normalizing dynamic values such as tokens, timestamps, or identifiers before computing these scores.

We compute the same comparison between the captured POST response and the no-query GET control. This produces four request-level classes:

\begin{itemize}
    \item \textbf{A: Strong interchangeability}: the converted GET matches the POST, but the no-query GET does not.
    \item \textbf{B: Ambiguous}: the converted GET matches the POST, and the no-query GET also matches the POST.
    \item \textbf{C: Baseline collision}: the converted GET does not match the POST, but the no-query GET does.
    \item \textbf{D: Not interchangeable}: neither GET response matches the POST.
\end{itemize}

The main prevalence measurement uses only class A as high-confidence evidence of method interchangeability. Classes A and B together are reported as a broader upper bound, since B may include generic responses or endpoints that ignore parameters. We also label route reachability separately: a GET response with status 2xx, 3xx, or 400 is treated as reaching the route, while 401, 403, 404, 405, 415, 422, 5xx, timeout, and network-error outcomes are treated as rejected. A 400 response can indicate that the GET reached the same handler, but it is not counted as acceptance unless the response is equivalent.

We manually validate the classifier after replay. This validation is not used to define the headline class-A measurement. Instead, it checks the reliability of the detector through manual inspection of sampled positive and negative cases.

\subsection{Design Rationale}

The methodology is intentionally conservative. The crawler uses real browser execution to capture requests generated by modern client-side applications, but replay is deterministic and recorded in JSONL files for later analysis. The no-query GET baseline prevents generic endpoint responses from being counted as strong evidence. Finally, response equivalence is treated as a detection criterion rather than proof of exploitability: a high-confidence interchangeable request shows that POST-like input can be represented as a GET URL with an equivalent observed response, but any security consequence depends on the endpoint semantics and surrounding defenses. This answers~\ref{Q2}, showing how \ac{MIDE} automatically detects method interchangeability at scale reliably, avoiding false positives from generic responses.

\section{Experiment}
\label{sec:experiment}

We evaluate our methodology on the Tranco top 10k~\cite{tranco}.\footnote{We use the Tranco list generated on 07 May 2026, available at \url{https://tranco-list.eu/list/VQWQN}.} The large-scale crawl was performed without authentication. The crawler performed the interaction-driven exploration described in Section~\ref{sec:methodology}, recording internal-domain and subdomain POST requests generated by page loads, forms, and selected button interactions.

For every collected POST request, we applied \ac{MIDE}. Supported bodies were transformed into GET candidates using the conversion rules in Section~\ref{sec:detection}. If conversion failed, the replay was skipped for that request. For successfully converted requests, we replayed the converted GET and a no-query baseline GET, then compared both responses against the captured POST response.

\subsection{Analysis Population}

Table~\ref{tab:analysis-coverage} reports the population that reaches each stage of the analysis. Of the 10,000 sites, 3,200 produced captured POST requests, for a total of 21,280 captured POST request records, and 2,734 had at least one POST that could be converted and replayed as a GET. Of the captured records, 15,639 (73.5\%) used one of the three supported body formats, while 14,244 (66.9\%) were converted into GET candidates. 5,641 (26.5\%) had unknown or unsupported content types. Because the crawler uses a strong denylist for potentially destructive or state-changing actions and file uploads, the captured POST count should be read as a lower bound on the POST requests reachable by our measurement pipeline, not as the complete POST surface of these sites. Request-level prevalence is measured over 14,244 replayed POST request records, while route-level prevalence is measured over 8,291 unique replayed POST routes. A route is defined as scheme, host, path, normalized method, and normalized content type.

\begin{table}[t]
    \caption{Results of our unauthenticated large-scale analysis over the Tranco top 10k sites.}
    \label{tab:analysis-coverage}
    \centering
    \begin{tabular}{lr}
    \toprule

    \textbf{Class} & \textbf{Count} \\

    \midrule

    Sites with captured POSTs       & 3,200   \\
    Sites with replayed POSTs       & 2,734   \\
    Captured POST requests          & 21,280  \\
    Replayed POST requests          & 14,244  \\
    Conversion-skipped POSTs        & 7,036   \\
    Unique replayed POST routes     & 8,291   \\

    \bottomrule
    \end{tabular}
\end{table}

We find 407 sites with at least one high-confidence interchangeable request, 4.1\% of all tested sites, 12.7\% of sites with captured POSTs, and 14.9\% of sites with replayed POSTs.

\subsection{Baseline-Control Classes}

We treat the no-query GET as a baseline, separating converted-parameter equivalence from responses that are equivalent regardless of parameters. Table~\ref{tab:baseline-classes} shows the resulting request-level classes among replayed POSTs. Class A is the high-confidence class: the converted GET matches the POST, but the no-query GET does not. Class B is an upper-bound class; here, both GETs match the POST. Class C is a baseline collision and is not treated as interchangeable. Class D covers the remaining non-interchangeable requests.

\begin{table}[t]
    \caption{Request-level baseline-control classes among replayed POSTs.}
    \label{tab:baseline-classes}
    \centering
    \begin{tabular}{lr}
    \toprule

    \textbf{Class} & \textbf{Requests} \\

    \midrule

    A: Strong interchangeability    & 1,228  (8.6\%)    \\
    B: Ambiguous                    & 1,782  (12.5\%)   \\
    C: Baseline collision           & 242    (1.7\%)    \\
    D: Not interchangeable          & 10,992 (77.2\%)   \\

    \bottomrule
    \end{tabular}
\end{table}

The baseline control enables us to rule out a large number of apparent equivalences not due to method interchangeability. Among the 3,010 requests in classes A and B, 1,782 (59.2\%) are class B ambiguous cases, where the no-query GET matches the POST response as well as the converted GET. These cases are not treated as high-confidence interchangeable requests because the parameters may not affect the response, but they would appear interchangeable without the baseline control. Class A contains 1,228 requests (40.8\%) that are high-confidence interchangeable requests.

\subsection{Comparator Validation and Sensitivity}

We validate the response-equivalence detector with a manual review of 60 sampled replay records: 30 class-A positives and 30 class-D negatives. The reviewer inspected the original POST, the converted GET, the no-query GET control, the request parameters, comparator scores, and response diffs, then labeled whether the compared responses had the same application-level outcome. All 30 class-A converted GET responses were manually judged equivalent to their POST responses, giving an observed class-A precision of 100\%. None of the 30 usable class-D records was judged equivalent to the POST response. This corresponds to an observed false-negative rate of 0\% among usable class-D annotations. The no-query baseline was also confirmed manually for all 30 class-A records; one class-A no-query response appeared equivalent on manual review. Overall, this manual validation supports the central assumption of the measurement: the comparator is highly accurate at distinguishing equivalent from non-equivalent responses in the sampled records.

\subsection{Prevalence}

Table~\ref{tab:main-prevalence} reports the main prevalence numbers using class A only on three levels, each answering a different question.
Site-level prevalence shows sites that have at least one high-confidence interchangeable request. Route-level prevalence presents how many distinct replayed POST routes are affected. Request-level prevalence asks how many individual replayed POST instances are affected.

\begin{table}[t]
    \caption{High-confidence method interchangeability prevalence.}
    \label{tab:main-prevalence}
    \centering
    \begin{tabular}{lrr}
    \toprule

    \textbf{Population}    & \textbf{Interchangeable} & \textbf{Total} \\

    \midrule

    All tested sites          & 407   (4.1\%)  & 10,000  \\
    Sites with captured POSTs & 407  (12.7\%)  & 3,200   \\
    Sites with replayed POSTs & 407  (14.9\%)  & 2,734   \\
    Replayed routes           & 646    (7.8\%) & 8,291   \\
    Replayed requests         & 1,228  (8.6\%) & 14,244  \\

    \bottomrule
    \end{tabular}
\end{table}

\subsection{Conversion and Replay Failures}

Next, we analyze and explain what the system misses before prevalence is computed. Table~\ref{tab:conversion-content-type} reports conversion success, replay success, and converted-response equivalence by content type. In this table, equivalence means that the converted GET matched the POST response; it is independent of the no-query baseline and therefore includes both A and B cases.

\begin{table}[t]
    \caption{Conversion, replay, and broad equivalence by content type.}
    \label{tab:conversion-content-type}
    \centering
    \small
    \setlength{\tabcolsep}{3pt}
    \begin{tabular}{@{}lrrrr@{}}
    \toprule

    \textbf{Content type} & \textbf{Captured} & \textbf{Converted} & \textbf{Replayed} & \textbf{Equivalent} \\

    \midrule

    Form-urlencoded & 7,310 & 7,310 & 7,182 & 2,117     \\
    JSON            & 7,570 & 6,231 & 6,139 & 692       \\
    Multipart       & 759   & 703   & 702   & 201       \\
    Unknown         & 5,641 & 0     & 0     & 0         \\

    \bottomrule
    \end{tabular}
\end{table}

Form-urlencoded bodies converted cleanly, with 7,310 of 7,310 records converted and 7,182 successful replays. JSON is more fragile: 6,231 of 7,570 JSON records converted (82.3\%), and 6,139 of those replayed successfully (98.5\%). Multipart conversion succeeded for 703 of 759 records (92.6\%), with 702 successful replays. All captured multipart bodies were fields-only; therefore, the lossy multipart confirmation population, where files are removed from the GET replay, was zero. JSON conversion produced 1,389 lossy conversions, and 29 of the equivalent JSON replays were lossy. Appendix~\ref{app:miss-reasons} provides a detailed breakdown of conversion and replay failures, non-equivalent replays, and practical constraints such as query length limits.

\subsection{Endpoint Taxonomy}

We next classify captured endpoints to understand what kinds of functionality are affected. The classifier is a lightweight heuristic, and uses replay record signals: host and path tokens, request parameter names, response title, response-body snippets, and normalized content type. We refer to its output as potentially security-relevant endpoint labels rather than confirmed functionality. These should therefore be read as broad categories, interpreted together with the observed HTTP status and replay outcome, rather than as precise classifications or evidence of successful authenticated access.

Among all captured POST requests, the most common labels were search or filtering (5,104 requests, 24.0\%), analytics or tracking (3,421 requests, 16.1\%), API or RPC endpoints (3,043 requests, 14.3\%), and authentication (2,188 requests, 10.3\%); 2,218 (10.4\%) endpoints remain unknown. Among high-confidence interchangeable requests, the largest labels were search or filtering (379 requests, 30.9\%), analytics or tracking (186 requests, 15.1\%), API or RPC (121 requests, 9.9\%), and authentication (109 requests, 8.9\%); 94 requests (7.7\%) remain unknown. The fine-grained potentially security-relevant labels are authentication, account settings, profile updates, password or email changes, cart or checkout flows, content creation, and admin or dashboard endpoints. We group newsletter or marketing, search or filtering, contact or feedback, and analytics or tracking as likely lower-sensitivity labels, and keep API or RPC and unknown labels separate because their semantics are ambiguous from the available signals. This grouping is necessarily coarse, as a search or filtering endpoint may enable XS-Search when its results depend on authenticated or personalized data and are observable cross-origin. However, our unauthenticated measurement cannot establish such user-dependent leakage, which would require a separate analysis of cross-origin observability and attacker capabilities.

\begin{table}[t]
    \caption{Potentially security-relevant endpoint labels among 14,244 replayed and 1,228 interchangeable POST requests, based on the endpoint taxonomy heuristic.}
    \label{tab:security-surface}
    \centering
    \begin{tabular}{lrr}
    \toprule

    \textbf{Label group} & \textbf{All} & \textbf{A only} \\

    \midrule

    Potentially security-relevant & 4,629 (32.5\%)    & 387   (31.5\%) \\
    Likely low-sensitivity     & 6,564 (46.1\%)    & 626   (51.0\%) \\
    API or RPC              & 1,754 (12.3\%)    & 121   (9.9\%)  \\
    Unknown                 & 1,297 (9.1\%)     & 94    (7.7\%)  \\

    \bottomrule
    \end{tabular}
\end{table}

Table~\ref{tab:security-surface} summarizes the potentially security-relevant endpoint labels over replayed POSTs and high-confidence interchangeable requests, answering~\ref{Q3}. We quantify how often method interchangeability appears in the measured population and which broad endpoint labels are affected. In the next section, we analyze the security implications of these results.

\section{Security Implications of HTTP Method Interchangeability}
\label{sec:security}

In this section, we analyze the security implications of method interchangeability for defenses that rely on HTTP method semantics. We focus on \ac{CSRF}, where applications often validate unsafe methods such as POST while treating GET as safe, and \ac{WCD}. Our measurements identify method-scoped defense gaps and cacheability conditions. While they do not prove state changes, sensitive-data exposure, or end-to-end exploitability, they show that method interchangeability can undermine assumptions made by defenses and caching layers.

\subsection{CSRF Token Validation Gaps}
\label{sec:csrf}

Modern \ac{CSRF} defenses frequently assume that state-changing operations use unsafe methods. This assumption appears both in server-side token checks on POST requests and in browser-side mitigations such as SameSite cookies, where cross-site top-level GET navigations may still carry cookies. Method interchangeability weakens this boundary: if a POST can be represented as an equivalent GET URL, the GET form may be easier to trigger cross-site and may not enforce the same token validation.

Therefore, we performed a targeted token-replay analysis over the high-confidence interchangeable requests. Starting from the class-A population from the previous section, we selected candidates whose captured POST contained a CSRF-like token name in the body or headers. We use the following case-insensitive token indicators: \texttt{csrf}, \texttt{xsrf}, \texttt{\_token}, \texttt{authenticity\_token}, \texttt{nonce}, \texttt{state}, \texttt{requestVerificationToken}, and \texttt{antiForgery}. This produced 115 token-bearing candidates across 46 sites. For each candidate, we replayed the original POST twice: with the token removed and changed, and replayed the converted GET with the original, removed, and changed token. We then compared each response with the first original POST replay.

We classify the results based on two dimensions: whether the POST accepts invalid tokens, and whether the GET accepts invalid tokens. We have the strongest evidence of a method-specific token-validation gap when the POST rejects invalid tokens, but the GET accepts them. In this case, the token check is effective on POST, but the interchanged GET form does not enforce the same check. If both POST and GET accept invalid tokens, the experiment does not show a GET-specific gap; it merely shows that token enforcement was not observed in that run. Similarly, if the POST baseline fails, we cannot assess whether POST-side CSRF validation is enforced.

\begin{table}[t]
    \caption{\ac{CSRF} token replay outcomes for 115 high-confidence interchangeable requests with CSRF-like tokens across 46 sites.}
    \label{tab:csrf-token-replay}
    \centering
    \begin{tabular}{lr}
    \toprule

    \textbf{Outcome}              & \textbf{Count} \\

    \midrule

    POST rejects; GET accepts   & 7  (6.1\%)    \\
    Both accept invalid tokens  & 78 (67.8\%)   \\
    Mixed, no GET-specific gap  & 3  (2.6\%)    \\
    POST baseline failed        & 16 (13.9\%)   \\
    GET validates tokens        & 1  (0.9\%)    \\
    Replay inconclusive         & 10 (8.7\%)    \\

    \bottomrule
    \end{tabular}
\end{table}

Table~\ref{tab:csrf-token-replay} reports the results. Seven candidates across two sites exhibit a pairwise method gap: the original POST succeeded, POST variants with missing or changed tokens were rejected, but the corresponding converted GET variants remained equivalent to the successful POST response. 78 candidates across 25 sites accepted invalid tokens for both POST and GET, so they do not show a method-specific gap. Three candidates showed mixed invalid-token behavior with no specific GET/POST difference, 16 had a failed POST baseline, one rejected invalid tokens for GET, and 10 were inconclusive.

The different behavior observed between POST and GET provides evidence of a method-specific token-validation gap enabled by method interchangeability. POST requests are protected by token checks that reject invalid tokens, while GET requests are accepted without valid tokens. We did not manually verify state change, user-specific sensitivity, or cross-site triggerability for these endpoints, so we report them as evidence of method-scoped defense gaps, rather than confirmed end-to-end \ac{CSRF} exploits.

These results do not imply that every token-bearing interchangeable endpoint is exploitable, but they demonstrate a measurable failure mode: a defense that is effective on POST may not be applied when the same operation is exposed through an equivalent GET.

\subsection{WCD Cacheability Conditions}
\label{sec:wcd}

Next, we test whether method-interchangeable endpoints exhibit cacheability behavior relevant to \ac{WCD}. \ac{WCD} occurs when dynamic content is cached under a URL that a cache treats as static, making the cached response retrievable by a request to the same URL~\cite{wcd_blog,wcd_paper,mirheidari2022web,golinelli2024hidden}. In our setting, we investigate whether responses that are originally reachable through POST become cacheable once method interchangeability exposes them through GET.

We run this experiment over the high-confidence interchangeable request instances. Each request instance corresponds to one captured POST whose converted GET belongs to class A; this is distinct from the route-level count of 646 unique routes reported in Section~\ref{sec:experiment}. We follow Mirheidari et al.~\cite{mirheidari2022web}, using the same payloads and \texttt{.css} static extension. Each request instance is tested with 12 \ac{WCD} payloads designed to reach the same dynamic response surface while being treated as static by caches. For each generated attack URL, we send two requests. The first uses the GET headers from the converted request to populate any intermediary cache. The second is issued from a separate session with credentials removed to check whether the response is served from cache or from the origin. We treat a payload as a \ac{WCD}-style cache hit when the payloaded URL yields the same dynamic response, the second response is a cache hit, and the two bodies match. Some payloads do not reach the dynamic handler and instead return static errors or unrelated responses. Others route correctly but do not result in cache hits. A small number produce cache hits with different bodies, which we treat as insufficient to confirm payload-induced cacheability.

\begin{table}[t]
    \caption{\ac{WCD}-style cacheability outcomes over 1,228 high-confidence interchangeable request instances. Request instances are counted once; the site column counts sites with at least one request instance in each outcome.}
    \label{tab:wcd-confirmation}
    \centering
    \footnotesize
    \begin{tabular}{@{}lrr@{}}
    \toprule

    \textbf{Outcome} & \textbf{Request Instances} & \textbf{Sites} \\

    \midrule

    Payload-induced cache hit       & 56  (4.6\%)  &  9   (2.2\%)     \\
    Normal URL already cacheable    & 136 (11.1\%) &  50  (12.3\%)    \\
    Cache HIT, body mismatch        & 5   (0.4\%)  &  3   (0.7\%)     \\
    Routed, no cache HIT            & 338 (27.5\%) &  133 (32.7\%)    \\
    Payload not routed/static error & 685 (55.8\%) &  262 (64.4\%)    \\
    Not dynamic                     & 8   (0.7\%)  &  1   (0.2\%)     \\

    \bottomrule
    \end{tabular}
\end{table}

Table~\ref{tab:wcd-confirmation} presents the results. We tested 1,228 high-confidence interchangeable request instances across 407 sites, with 12 \ac{WCD} payloads per request instance. Each request instance is counted once according to the strongest outcome observed across all payloads. The site column is non-exclusive: it counts sites with at least one request instance in the corresponding outcome, so site shares do not sum to 100\%. Overall, 56 request instances (4.6\%) across 9 sites (2.2\%) had at least one \ac{WCD} payload-induced cache hit. Another 136 request instances (11.1\%) across 50 sites (12.3\%) were already cacheable at the original converted GET URL, without requiring a payload. This is distinct from payload-induced \ac{WCD}, but still indicates that method interchangeability can expose dynamic responses through cacheable GET URLs. A further 685 request instances did not route any payload to the dynamic handler or produced static-looking errors, and 338 routed at least one payload but did not show evidence of cache hits.

These results show that method interchangeability can turn a POST-equivalent dynamic response into a cacheable GET resource. Together with the \ac{CSRF} token-replay analysis, this answers~\ref{Q4} by showing how method interchangeability can weaken defenses that rely on HTTP method semantics.

\section{Discussion}

Our results show that HTTP method interchangeability is not an isolated artifact. In the Tranco top 10k, 407 of 10,000 tested sites (4.1\%), 407 of 3,200 sites with captured POSTs (12.7\%), and 407 of 2,734 sites with replayed POSTs (14.9\%) contained at least one high-confidence interchangeable request. At the route and request levels, the corresponding rates were 646 of 8,291 replayed routes (7.8\%) and 1,228 of 14,244 replayed requests (8.6\%). A behavior often treated as a corner case, therefore appears across hundreds of popular sites and thousands of request instances.

Many endpoints return equivalent responses even when the converted parameters are removed; we classify these as class-B cases. In our measurement, 1,782 replayed requests were ambiguous class-B cases, compared with 1,228 high-confidence class-A cases. This filters out many false positives, but also suggests that interchangeability may be more widespread than our high-confidence estimate.

Our endpoint taxonomy analysis also assigns potentially security-relevant labels to some interchangeable requests, not only search boxes or analytics beacons. Among high-confidence interchangeable requests, 387 (31.5\%) receive a potentially security-relevant label, and 121 (9.9\%) are API or RPC endpoints. These labels do not show that the crawler accessed authenticated functionality; they may correspond to public routes, login redirects, authorization errors, generic API endpoints, or lexical misclassifications. Although the taxonomy is heuristic, the fine-grained labels identify areas for follow-up where method semantics may matter, including authentication, profile update, and account settings.

We also examine the security relevance of interchangeability. Among 115 token-bearing high-confidence interchangeable requests, seven cases across two sites showed a method-specific \ac{CSRF} gap: invalid tokens were rejected for POST but accepted for the equivalent GET. Across 1,228 high-confidence interchangeable request instances on 407 sites, we observe \ac{WCD}-style payload-induced cache hits on 56 request instances across 9 sites, and 136 request instances across 50 sites were already cacheable at the original converted GET URL. Method interchangeability can therefore expose both method-scoped token-validation gaps and cacheable POST-equivalent responses.

\subsection*{Limitations}

Our crawler uses a strong denylist for potentially destructive or state and session-changing actions and file uploads. This reduces the risk of side effects on live sites, but it means that the POST population we collect is a lower bound.

The measurement is unauthenticated and therefore does not cover functionality available only after login. Coverage is further bounded by the 10-page and depth-5 crawl limits and the caps of 5 forms and 10 standalone buttons per page. The crawler triggers selected form and button interactions rather than exhaustively exploring all event types and UI states, while its benign synthetic inputs may fail application-specific validation or miss behaviors that depend on realistic user data. Consequently, the results characterize only the POST surface reached by this interaction strategy.

Browser automation introduces additional sources of undercoverage. Sites may behave differently in a headless browser or restrict the crawler through bot detection, despite the use of browser-stealth measures. More generally, Roth et al. show that client characteristics such as the user agent, network access method, and language settings can produce inconsistent client-side security policies~\cite{roth2022securitylottery}. Such behavior can prevent pages, controls, or requests from being observed and may bias the measured population toward sites that permit automated access.

Finally, replay is limited to supported body formats and cannot always preserve their semantics when mapping them to a query string. Unsupported content types are skipped, while nested JSON values, nulls, and multipart file fields can make conversion fail or become lossy. Converted requests that exceed the configured 8,192-byte query-length limit are also excluded, and servers and intermediaries may enforce different URL limits. These constraints make the reported prevalence a conservative estimate for the explored population.

\subsection{Conclusion}

This paper presented \ac{MIDE}, a browser-based pipeline for measuring HTTP method interchangeability through POST-to-GET rewriting and response-equivalence analysis. We release its code and experimental configuration as open source to assist practitioners and researchers in detecting method interchangeability. We use \ac{MIDE} on the Tranco top 10k, in the first large-scale measurement of HTTP method interchangeability in the wild. We discover that high-confidence method interchangeability is observable at scale, affecting 4.1\% of all tested sites, 12.7\% of sites with captured POSTs, and 14.9\% of sites with replayed POSTs. The security analysis further shows why this behavior matters: equivalent GET requests can expose method-scoped token-validation gaps, and POST-equivalent dynamic responses can become cacheable GET resources.

\section*{Ethical Considerations}

Our measurement pipeline is designed to minimize the impact on the sites we visit. In the Tranco run, crawling was bounded to 10 pages and depth 5, used a 100 captured-POST stop threshold per site, and the batch runner processed sites sequentially with a 5-second delay between sites. Within each page, the crawler interacted with at most 5 forms and 10 standalone buttons, used benign synthetic inputs, skipped file inputs, and never injected malicious payloads.

We also include safeguards to avoid actions that are likely to be destructive or sensitive. The crawler avoids potentially destructive or state and session-changing action keywords in button text and captured POST URL paths, filters denylisted URLs, stays within the target site, and drops multipart requests that contain file-upload fields. The main replay stage sends only converted GET requests and rate-limits them with a 2-second delay between replayed records.

\section*{LLM Usage Statement}

Anthropic's Claude Sonnet 4 and OpenAI models assisted with code completion, boilerplate, and implementation details. The authors manually reviewed, tested, and adapted all outputs. LLMs were used for editorial purposes in this manuscript, and all outputs were inspected by the authors to ensure accuracy and originality. The research design, methodology, analysis, and conclusions were created by the authors.

\section*{Acknowledgments}

This paper was partly supported by the EU Digital Europe Programme under GA 101123118 (SPECTRO).

\bibliographystyle{IEEEtran}
\bibliography{main}

\appendices

\section{Detailed Web Framework Analysis}
\label{app:framework-details}

\subsection{Unified Parameter Access}

Several frameworks expose abstractions that partially or fully unify parameter sources. \textbf{Flask} (Python) provides \texttt{request.values}, which combines \texttt{request.args} and \texttt{request.form}; JSON remains separate via \texttt{request.get\_json()}.\footnote{Flask 3.x, \url{https://flask.palletsprojects.com/en/stable/api/\#flask.Request.values}} \textbf{Rails} (Ruby) exposes a unified \texttt{params} object that merges query parameters, form data, route parameters, and parsed JSON payloads.\footnote{Rails 8.x, \url{https://guides.rubyonrails.org/api_app.html}} \textbf{Sinatra} follows a similar model for route and request parameters, although its documentation is less explicit for structured bodies such as JSON.\footnote{Sinatra 4.x, \url{https://sinatrarb.com/intro.html}}

\textbf{Vanilla PHP} provides the most permissive model through \texttt{\$\_REQUEST}, which merges \texttt{\$\_GET} and \texttt{\$\_POST} (and optionally cookies, depending on configuration). \textbf{Laravel} builds on this with accessors such as \texttt{input()} and \texttt{all()}, which merge query parameters, form data, and JSON payloads.\footnote{PHP 8.x, \url{https://www.php.net/manual/en/reserved.variables.request.php}; Laravel, \url{https://laravel.com/docs/requests}}

\textbf{Spring Boot} (Servlet-based Spring MVC), \textbf{ASP.NET Core}, and \textbf{Gin} provide partial or adaptive unification rather than a single merged namespace. In Spring MVC, \texttt{@RequestParam} binds Servlet request parameters, i.e., query parameters and form data, while JSON remains separate via \texttt{@RequestBody}.\footnote{Spring Framework, \url{https://docs.spring.io/spring-framework/reference/web/webmvc/mvc-controller/ann-methods/requestparam.html}} In ASP.NET Core, model binding can draw from form fields, the request body, route data, query string parameters, and uploaded files, but source selection remains explicit through attributes such as \texttt{[FromQuery]}, \texttt{[FromForm]}, and \texttt{[FromBody]}.\footnote{ASP.NET Core 10.0, \url{https://learn.microsoft.com/en-us/aspnet/core/mvc/models/model-binding?view=aspnetcore-10.0}} In Gin, Go's \texttt{Request.Form} merges query parameters with form-encoded body data after form parsing, and \texttt{ShouldBind} uses query binding on GET and body/form binding on POST depending on content type.\footnote{Gin documentation, \url{https://gin-gonic.com/en/docs/binding/bind-query-or-post/}; Go \texttt{net/http}, \url{https://pkg.go.dev/net/http}}

By contrast, \textbf{Django}, \textbf{Express}, \textbf{FastAPI}, \textbf{NestJS}, \textbf{Next.js}, and \textbf{SvelteKit} keep query parameters and request bodies more explicit. Django exposes \texttt{request.GET} and \texttt{request.POST} separately and provides explicit method-restriction decorators. Express separates \texttt{req.query}, \texttt{req.body}, and \texttt{req.params}, and removed legacy unified accessors such as \texttt{req.param()} in recent versions. FastAPI enforces explicit parameter-source declarations. NestJS uses decorators such as \texttt{@Query()} and \texttt{@Body()}; Next.js uses method-specific handlers; and SvelteKit uses method-specific actions and separates URL parameters from request bodies. These frameworks require deliberate developer choices to approximate method interchangeability and do not enable it through default abstractions.

\subsection{Routing and Method Constraints}

Unified parameter access alone does not imply method interchange. The decisive factor is whether the same endpoint is reachable through multiple HTTP methods.

\textbf{Vanilla PHP} is the clearest default case: scripts are generally reachable through any method unless the application explicitly checks \texttt{\$\_SERVER['REQUEST\_METHOD']}. Combined with \texttt{\$\_REQUEST}, this yields both method-agnostic reachability and unified parameter access.

\textbf{Spring Boot} and \textbf{ASP.NET Core} require a more careful qualification. In Spring MVC, generic mappings such as \texttt{@RequestMapping} may accept multiple HTTP methods unless narrowed with \texttt{@GetMapping} or \texttt{@PostMapping}. Similarly, in ASP.NET Core, a generic \texttt{[Route]} attribute on an action accepts all HTTP methods unless a verb-specific attribute is used.\footnote{Spring Framework current reference, \url{https://docs.spring.io/spring-framework/reference/web/webmvc/mvc-controller/ann-requestmapping.html}; ASP.NET Core 10.0, \url{https://learn.microsoft.com/en-us/aspnet/core/mvc/controllers/routing?view=aspnetcore-10.0}} In both frameworks, method interchange is therefore configuration-dependent and is strongest for query and form-style inputs, not for JSON bodies.

\textbf{Django} keeps \texttt{request.GET}, \texttt{request.POST}, and \texttt{request.body} separate, but plain views are reachable through multiple HTTP methods unless the application checks \texttt{request.method} or uses explicit decorators such as \texttt{require\_http\_methods}.\footnote{Django 6.0, \url{https://docs.djangoproject.com/en/6.0/ref/request-response/}} By default, Django is therefore method-permissive, but source-explicit.

\textbf{Laravel}, \textbf{Flask}, \textbf{Rails}, \textbf{Sinatra}, and \textbf{Gin} require explicit developer action to enable multi-method reachability, e.g., \texttt{Route::match}, \texttt{methods=['GET','POST']}, \texttt{via: [:get, :post]}, or separate route registration. These frameworks make inputs look similar once a handler is shared, but do not enable method interchange by default.

\subsection{Array Parameter Parsing}

Beyond route reachability, frameworks differ significantly in how they parse array-like query parameters, which affects the correctness of request transformation.

\textbf{Spring Boot} accepts repeated keys (e.g., \texttt{ids=1\&ids=2}) but does not natively support PHP-style array syntax or indexed notation. \textbf{PHP} supports PHP-style arrays (e.g., \texttt{ids[]=1}) and indexed arrays, but not repeated keys in the same way. \textbf{Flask} supports repeated keys. \textbf{Rails} and \textbf{Sinatra} are the most permissive, supporting multiple encodings, though indexed arrays may be interpreted as maps rather than true arrays.

These differences inform our design of automated detection, as POST-to-GET transformations must preserve semantic equivalence under framework-specific parsing rules.

\newpage
\balance
\section{Reasons for Missed Replays and Equivalences}
\label{app:miss-reasons}

\begin{table}[t]
    \caption{Priority-ordered primary replay labels. Percentages are out of the 21,280 captured POST requests.}
    \label{tab:miss-reasons}
    \centering
    \begin{tabular}{lrr}
    \toprule

    \textbf{Reason} & \textbf{Requests} \\

    \midrule

    Replayed but not equivalent         & 10,414 (48.9\%)   \\
    Unsupported content type            & 5,641 (26.5\%)    \\
    Clean equivalent replay candidate   & 2,887 (13.6\%)    \\
    Body parse error                    & 838   (3.9\%)     \\
    Empty body                          & 664   (3.1\%)     \\
    Oversized query after conversion    & 465   (2.2\%)     \\
    Network or replay error             & 193   (0.9\%)     \\
    Top-level JSON array failure        & 178   (0.8\%)     \\

    \bottomrule
    \end{tabular}
\end{table}

Table~\ref{tab:miss-reasons} categorizes conversion and replay outcomes using a priority-ordered primary label. 10,414 captured POSTs (48.9\%) were converted and replayed but were not equivalent to the POST response. In total, 3,010 replayed POSTs produced an equivalent converted GET response before applying the no-query baseline, corresponding to classes A and B in Table~\ref{tab:baseline-classes}. Of these, 2,887 captured POSTs (13.6\%) are clean equivalent replay candidates. The remaining 123 equivalent responses receive earlier labels: 109 had empty POST bodies and 14 exceeded the configured 8,192-byte query threshold after conversion.
The dominant pre-replay limitation is unsupported content type, which accounts for 5,641 captured POSTs (26.5\%). Query length is another practical constraint: after conversion, the 95th percentile query length was 3,536 bytes for form-urlencoded bodies, 6,133 bytes for JSON, and 5,372 bytes for multipart bodies; 197 form-urlencoded, 255 JSON, and 13 multipart conversions exceeded the configured 8,192-byte threshold.

The JSON shape analysis, computed over the 7,570 captured JSON POST records with non-exclusive labels, explains part of the fragility. Most JSON records were top-level objects (6,783 records, 89.6\%), but nested objects appeared in 4,548 records (60.1\%) and nested arrays in 2,157 records (28.5\%). Top-level arrays were less common (436 records, 5.8\%), mixed arrays appeared in 22 records (0.3\%), and scalar top-level JSON was rare (5 records, 0.1\%). Body parse errors appeared in 346 JSON records (4.6\%).

\end{document}